\documentclass[useAMS, usenatbib]{mn2e}
\usepackage{amsmath}
\usepackage{graphicx}
\usepackage{epstopdf}
\usepackage{natbib}

\def\be{\begin{equation}}
\def\ee{\end{equation}}
\def\ba{\begin{eqnarray}}
\def\ea{\end{eqnarray}}
\def\sl{\mathrm{sl}}

\def\rhomax{\rho_{\mathrm{max}}}
\def\go{\mathrel{\raise.3ex\hbox{$>$}\mkern-14mu
             \lower0.6ex\hbox{$\sim$}}}

\def\lo{\mathrel{\raise.3ex\hbox{$<$}\mkern-14mu
             \lower0.6ex\hbox{$\sim$}}}

\begin{document}
\title{Analytical Model of Tidal Distortion and Dissipation for a Giant Planet with a Viscoelastic Core}
\author[Natalia I Storch and Dong Lai]
{Natalia I Storch\thanks{Email: nis22@cornell.edu, dong@astro.cornell.edu}
and Dong Lai\\
Center for Space Research, Department of Astronomy, Cornell University, Ithaca,
NY 14853, USA\\}

\pagerange{\pageref{firstpage}--\pageref{lastpage}} \pubyear{2015}

\label{firstpage}
\maketitle

\begin{abstract}
We present analytical expressions for the tidal Love numbers of a
giant planet with a solid core and a fluid envelope.  We model the
core as a uniform, incompressible, elastic solid, and the envelope as
a non-viscous fluid satisfying the $n=1$ polytropic equation of state.
We discuss how the Love numbers depend on the size, density, and shear
modulus of the core.  We then model the core as a viscoelastic Maxwell
solid and compute the
tidal dissipation rate in the planet as characterized by the imaginary
part of the Love number $k_2$.  Our results improve upon existing
calculations based on planetary models with a solid core and a uniform
($n=0$) envelope.  Our analytical expressions for the Love numbers can
be applied to study tidal distortion and viscoelastic dissipation of
giant planets with solid cores of various rheological properties, and
our general method can be extended to study tidal distortion/dissipation of super-earths.
\end{abstract}


%%%%%%%%%%%%%%%%%%%%%%%%%%%%%%%%%%%%%%%%%%%%       
\section{Introduction}

Tidal effects play an important role in understanding many puzzles
associated with planet/exoplanet formation and evolution.  One example
involves high-eccentricity migration of giant planets: tidal dissipation in the
planet is responsible for circularizing the planet's orbit, leading to
the creation of hot Jupiters  (e.g. Wu \& Murray 2003, Fabrycky \& Tremaine 2007,
Correia et al. 2011, Naoz et al. 2012, Storch et al. 2014, Petrovich 2015).

The levels of tidal dissipation in giant planets suggested by both
Solar System (e.g. Goldreich \& Soter 1966, Yoder \& Peale 1981,
Lainey et al. 2009, 2012) and extrasolar (e.g. Socrates et al. 2012,
Storch \& Lai 2014) constraints cannot easily be explained by simple
viscous dissipation in the turbulent fluid envelope (Goldreich \&
Nicholson 1977). While several mechanisms based on wave excitation and
dissipation in the envelope (ocean) have been 
studied (e.g. Ogilvie \& Lin 2004, Ivanov \& Papaloizou 2007; see
Ogilvie 2014 for a review), it remains unclear whether they can
provide sufficient dissipation.

Dissipation in the solid cores of giants planets is another possible
source of dissipation. Previous works (Remus et al. 2012, Storch \&
Lai 2014, Remus et al. 2015) have employed analytical formulae for the
tidal dissipation in a two-layer planet consisting of a uniform,
incompressible, viscoelastic core and a uniform non-dissipative ocean
(e.g. Dermott 1979).
These works demonstrate that, while there are significant 
uncertainties in the rheologies of the solid core,
it is in principle possible for dissipation
in the core to be substantial enough to account for existing
constraints, particularly the dependence of dissipation on the tidal
forcing frequency (Storch \& Lai 2014).

The advantage of analytical models for the tidal
deformation lies in the use of the ``correspondence principle'', in which
the analytical formulae derived for the tidal deformation of an
elastic body may be generalized to a {\it visco}elastic body via
introduction of a complex shear modulus (Biot 1954). 
This allows various rheologies for the solid core to be employed 
in calculating the tidal dissipation.  
In this paper, we extend previous works by considering a fluid envelope
(ocean)
of non-uniform density, rather than a uniform one. In particular, we
show that if the ocean obeys the $n=1$ polytropic equation of state ($P\propto
\rho^2$), relatively simple analytical expressions for the tidal Love
numbers can be obtained.
The $n=1$ polytrope is appropriate for giant planets, as it correctly
reproduces the fact that their radii are nearly independent of their
masses.

In section 2, we set up the analytical problem of calculating the
tidal distortion in a two-layer giant planet. In section 3, we present
the solution for the tidal radial deformation of the core,
characterized by the Love number $h_{2c}$, and the change in the
self-gravity of the planet, characterized by the Love number $k_2$. In
section 4 we give several examples of the uses of these formulae. We
discuss our results and conclude in section 5.

\section{Setup and Schematic Solution}

We consider distortion of a planet by an $l=2$ tidal potential. Let
the perturber have mass $M_\star$ and be a distance $a$ away.  Let the
planet have mass $M_p$ and radius $R$, and possess a solid core of
radius $R_c$. We assume the core is incompressible, with uniform
density $\rho_c$ and shear modulus $\mu$. We model the planet's fluid
envelope as an $n=1$ polytrope, such that its pressure ($P$) and
density ($\rho$) profiles satisfy the relation
\be
P(r)=K \rho(r)^2,
\ee
where $K$ is a constant. 

\subsection{Equilibrium Structure}

In the absence of a perturber, the planet is in hydrostatic
equilibrium. The planet's gravitational potential $\Phi$ and pressure
profile $P$ satisfy the equations
\ba
\nabla^2\Phi &=& 4\pi G\rho, \\
\nabla P &=& -\rho \nabla\Phi. 
\ea
It follows that the density profile in the fluid ocean is given by 
\be
\rho(r) = \rho_0 {\sin[ q\,(1-r/R)]\over q\,r/R},
\label{rho0ofr}
\ee
where
\ba
q^2 &=& \frac{2\pi G R^2}{K}, \\
\rho_0 &=& {q^2 M_p\over 4 \pi R^3}.
\label{rho0}
\ea
For 
clarity, we define $\rhomax$ to be the
fluid density at $r=R_c+$ 
(just outside $R_c$):
\be
\rhomax \equiv \rho_0{\sin q(1-R_c/R)\over q R_c/R}.
\label{rhomax}
\ee
Since we demand the planet to be of mass $M_p$ and radius 
$R$, and the core to have radius $R_c$, this leads to a constraint
on the core-to-fluid density jump:
\be
{\rho_c\over\rhomax} = {3 R^2\over q^2 R_c^2}\left[{q R_c\over R} \cot q(1-R_c/R) + 1\right].
\label{rhocconstr}
\ee
In practice, 
for given $M_p$, $R$, $\rho_c$ and $R_c$ (or $M_c$), we solve for $q$
from equations (\ref{rho0})-(\ref{rhocconstr}).
For completeness, we give the
potential $\Phi$ inside the planet:
\be
{\small
\Phi(r) = \begin{cases} 
{2\over 3} \pi G\rho_c (r^2-R_c^2) - {GM_p\over R} \left[{\sin q(1-R_c/R)\over q R_c/R} + 1\right] \!\!\! &(r \le R_c) \\
-{GM_p\over R}\left[{\sin q(1-r/R)\over q r/R} + 1\right] & (R_c \le r \le R).
\end{cases}
\label{phi0ofr}}
\ee

\subsection{Tidal Perturbation}

We now turn on the $l=2$ tidal perturbation and calculate the
resulting deformation of the planet. We set up the problem such that
the $z$ axis joins the centers of the planet and the perturber. In
this way, the problem has azimuthal symmetry. The perturbing
tidal potential (assumed small) is given by
\be
\overline{U}(r,\theta) \equiv U(r)Y_{20}(\theta) =  - \sqrt{4\pi\over 5} {G M_\star\over a^3} r^2 Y_{20}(\theta),
\ee
Hence we can assume all perturbed quantities are proportional to
$Y_{20}$. The perturbed Poisson's equation is given by
\be
\nabla^2\overline{\delta\Phi} = 4\pi G\overline{\delta\rho},
\ee 
where $\overline{\delta X}\equiv\delta X(r) Y_{20}(\theta)$ indicates the Eulerian perturbation to the variable $X$, and the perturbed equation of hydrostatic equilibrium in the liquid layer of the planet is given by
\be
\nabla \overline{\delta P} = -\overline{\delta\rho}\nabla\Phi -\rho\nabla\overline{V},
\label{HSEpert}
\ee
where $\overline{V}\equiv V(r)Y_{20}(\theta)\equiv
\overline{\delta\Phi} + \overline{U}$, and $\rho(r)$ and $\Phi(r)$ are
the unperturbed density profile and unperturbed gravitational
potential as derived in the previous subsection. The transverse
component of Eq. (\ref{HSEpert}) is $\overline{\delta P} = -\rho
\overline{V}$, while the radial component is $(\overline{\delta P})' =
-\overline{\delta\rho}\Phi' - \rho\overline{V}'$, where $'$ stands for
$\partial/\partial r$. These imply that $\overline{\delta\rho} =
\left(\rho'/\Phi'\right)\overline{V}$.

Inside the core, and outside the planet, the perturbed Poisson equation reduces to 
$\nabla^2 \overline{\delta\Phi} = 0$,
and is easily solved, yielding $\delta\Phi = b_1 r^2$ inside the core 
and $\delta\Phi=b_4 r^{-3}$ outside the planet, with $b_1$ and $b_4$ as yet unknown constants. 
In the fluid envelope, the Poisson equation reduces to
\be
\nabla^2 \overline{V} = 
4\pi G \overline{\delta\rho}
=4 \pi G {\rho'\over\Phi'} \overline{V}.
\label{veq}
\ee

So far the equations in this subsection are general (valid for all
envelope equation of state). for the $n=1$ EOS, $P'=-\rho\Phi'$ gives
$\left(\rho'/\Phi'\right) = -1/(2K)$ and we have
\be
\nabla^2 \overline{V} = -\left({q\over R}\right)^2 \overline{V}.
\ee
This equation admits a standard solution in the form of spherical Bessel functions ($j_2$ and $y_2$): 
\be
V(r) = b_2\, j_2\!\left(q {r\over R}\right) + b_3 \,y_2\!\left(q {r\over R}\right)
\ee
inside the fluid envelope, with $b_2$ and $b_3$ constants to be determined. 

The unknown constants $b_1$, $b_2$, $b_3$, $b_4$ may now be solved for by matching boundary conditions at
$r=R$ and at $r=R_c$:
\ba
\delta\Phi(R_+) &=& \delta\Phi(R_-), \label{PhiR} \\
\delta\Phi(R_{c,+})&=&\delta\Phi(R_{c,-}), \\
\left(\delta\Phi'\right)_{R_-}&=&\left(\delta\Phi'\right)_{R_+}=-\frac{3}{R}\delta\Phi(R_+), \\
\left(\delta\Phi'+4\pi G\rho \xi_r\right)_{R_{c,-}}&=&\left(\delta\Phi'+4\pi G\rho \xi_r\right)_{R_{c,+}}.
\ea 
Since the fluid density vanishes at surface of planet, these introduce
only one additional unknown: the radial displacement at the core-fluid
interface, $\xi_r(R_c)$. Determining $\xi_r(R_c)$ requires solving for
the deformation of the core, matching the radial and transverse
tractions across the core-fluid interface. The procedure to follow is
similar to Love's classic solution for the deformation of an
incompressible, uniform, self-gravitating elastic body under an
external potential (Love 1911, Greff-Lefftz et al. 2005), with the
addition of an external pressure force due to the fluid envelope. We
find the final boundary condition is given by
\ba
{19\over 5}\mu\frac{\xi_r(R_c)}{R_c} &=& -\left(\frac{dP}{dr}+\rho_c g_c\right)\xi_r \label{traction}\\
&+& (\rhomax - \rho_c)(b_1 R_c^2 + U), \nonumber
\ea
where $g_c\equiv (4/3) \pi G \rho_c R_c$ is the gravitational
acceleration at $r=R_c$, and all the quantities on the RHS are
evaluated at $r=R_{c,+}$. Note that equation (\ref{traction}) is valid
for any envelope EOS. Together, the five boundary conditions, Eqs.~(\ref{PhiR})~--~(\ref{traction}), can
be solved for all the unknowns.

\section{Analytical Love Number Formulae}

\subsection{Non-dissipative tide}

We are interested in two dimensionless Love numbers. The first is the tidal Love number of the planet, defined as 
\be
k_2 \equiv \frac{\delta\Phi(R)}{U(R)}.
\ee
This specifies the magnitude of the quadrupole potential
produced by the distorted planet, $\overline{\delta\Phi} = k_2 U(R)
(R/r)^3 Y_{20}(\theta)$ (for $r > R$), and therefore determines the
effect of tidal distortion on the planet's orbit. The second is the
radial displacement Love number of the core, defined by
\be
h_{2c} \equiv -\frac{\xi_r(R_c)g_c}{U(R_c)}.
\ee
This specifies the shape of the inner core under the combined
influences of the external tidal field and the loading due to the
fluid envelope.

Following the schematic procedure outlined in the previous section, we find
\ba
h_{2c} &=& \frac{5}{q^2}\left(\frac{R}{R_c}\right)^3\left\{\alpha\left[1+\frac{2\bar{\mu}}{5\left(1-\frac{\rhomax}{\rho_c}\right)}\right]\right.  \label{h2c}\\
&& \left.\phantom{\frac{2\bar{\mu}}{5\left(1-\frac{\rhomax}{\rho_c}\right)}} - 3 \lambda \left(1-\frac{\rhomax}{\rho_c}\right)\right\}^{-1},\nonumber \\
k_2 &=& \frac{3h_{2c}}{q \alpha}\left(\frac{R_c}{R}\right)^2\left(1-\frac{\rhomax}{\rho_c}\right) + \frac{5 \gamma}{q \alpha} - 1,
\ea
where 
\be
\bar{\mu} \equiv 19\mu/(2\rho_c g_c R_c),
\ee 
\noindent and
\ba
\alpha &=& j_1(q)\left[\chi_c y_1\!\!\left(\chi_c\right) - 5 y_2\!\!\left(\chi_c\right)\right]
- y_1(q)\left[\chi_c j_1\!\!\left(\chi_c\right) - 5 j_2\!\!\left(\chi_c\right)\right], \nonumber\\
\lambda &=& y_1(q)j_2\!\!\left(\chi_c\right)-j_1(q)y_2\!\!\left(\chi_c\right), \label{lambda}\\
\gamma &=& j_2(q)\left[\chi_c y_1\!\!\left(\chi_c\right)-5 y_2\!\!\left(\chi_c\right)\right] 
- y_2(q)\left[\chi_c j_1\!\!\left(\chi_c\right)-5 j_2\!\!\left(\chi_c\right)\right],\nonumber 
\ea
with $\chi_c\equiv qR_c/R$.

\subsection{Dissipative tide}

We now consider the effects of viscous dissipation in the solid core.
According to the correspondence principle (Biot 1954), we may
generalize the calculation of tidal distortion of a non-dissipative
elastic core by adopting a complex shear modulus $\mu$, where the
imaginary part of $\mu$ accounts for dissipation in the 
{\it visco}elastic core. In general, the complex $\mu$ depends on the
tidal forcing frequency $\omega$ in the rest frame of the planet, and
its actual form depends on the rheology of the solid (e.g Henning et
al. 2009, Remus et al. 2012). Thus the complex $k_2 = k_2(\omega)$
also depends on the forcing frequency.

For example, assume the perturber is in a circular orbit with orbital
frequency $\Omega$ and the planet is spinning with frequency
$\Omega_s$. The forcing frequency is then $\omega =
2\Omega-2\Omega_s$. The torque on the planet and the energy transfer
rate from the orbit to the planet due to dissipation may be calculated
as (Storch \& Lai 2014)
\ba
T_z &=& \frac{3}{2}T_0\, {\rm Im}\!\left[k_2(2\Omega - 2\Omega_s)\right], \\
\dot{E} &=& \frac{3}{2}T_0\Omega\, {\rm Im}\!\left[k_2(2\Omega - 2\Omega_s)\right],
\ea
where $T_0 \equiv G(M_\star/a^3)^2 R^5$. See Storch \& Lai (2014) for
the more general case of a perturber on an eccentric orbit.

Note that $\dot{E}$ includes contributions both from dissipation into
heat, which occurs solely in the viscoelastic core, and from the
torque $T_z$ that acts to synchronize the rotation rate of the planet
with the orbital frequency of the perturber. Thus, the true tidal
heating rate received by the core is given by
\be
\dot{E}_{\rm heat} = \dot{E} - \Omega_s T_z.
\ee

\section{Applications of Love Number Formulae}

In this section we present several sample applications of the formulae
derived in the previous section. First we consider planets with
non-dissipative cores, then generalize to a complex shear modulus and
compute the tidal dissipation.

\subsection{Non-dissipative elastic core}

In Figure 1 we present the Love numbers for a giant planet (mass
$M_p=M_J$, radius $R=R_J$) with a core of constant density, $\rho_c =
6\,{\rm g\, cm^{-3}}$, but varying size, for several values of the
core shear modulus $\mu$. Our nominal reference value for $\mu$ is
that of undamaged rocky material at Earth-like pressures and
temperatures, $\mu_0 = 900$ kbar. Damaged rocky, or icy material can
have a lower shear modulus $\sim 40$ kbar (Goldsby \& Kohlstedt 2001,
Henning et al. 2009). However, little is known about both the
composition of giant cores and the behavior of rocky/icy materials
under high pressures and thus the value of $\mu$ is largely a free
parameter.
 
Based on Fig. 1, we note that the Love numbers generally behave as
expected. Cores with higher shear moduli are harder to deform,
resulting in smaller Love numbers. Planets with cores of larger radii
have more mass concentrated in the center, and thus $k_2$ decreases as
a function of $R_c$. At $R_c=0$, i.e. in the absence of a core, $k_2$
correctly defaults to the standard value for an $n=1$ envelope,
$(15/\pi^2) - 1$. At $R_c/R \approx 0.6$, the core mass is equal to
planet mass, i.e. the envelope has zero mass but still artificially
extends to $R=R_J$. In this case $k_2$ and $h_{2c}$ default to values
for a bare core:
\be
h_{2c,0} = \frac{5}{2(1+\bar{\mu})}, \quad ~~k_{2,0} =\frac{3}{2(1+\bar{\mu})}\left(\frac{R_c}{R}\right)^5.
\label{h0k0}
\ee

\begin{figure}
\scalebox{0.58}{\includegraphics{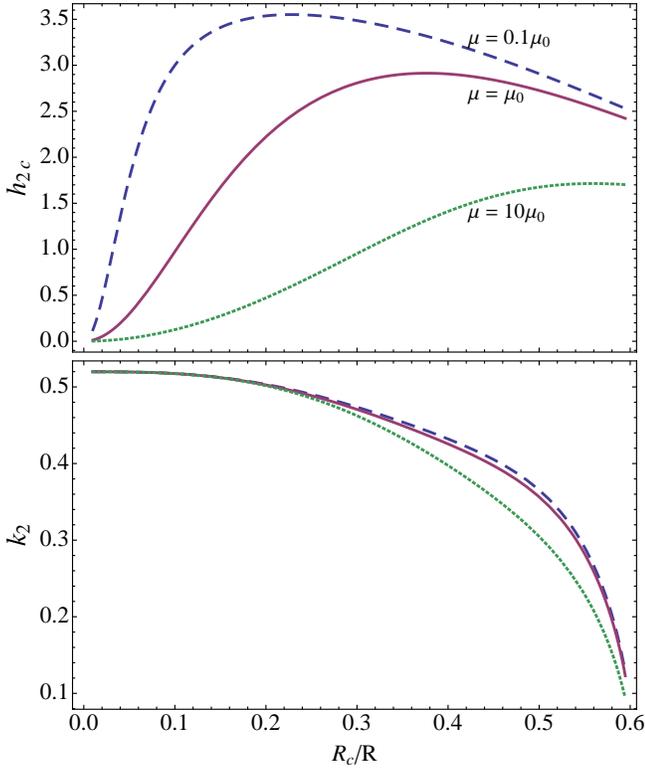}}
\caption{Sample tidal Love numbers $h_{2c}$ (top) and $k_2$ (bottom) for a gas giant with mass $M_J$, radius $R_J$, and core density $\rho_c=6\,{\rm g\,cm^{-3}}$, as a function of core radius for several values of core shear modulus, where $\mu_0=900$\,kbar.}
\end{figure}

In Figure 2 we present the Love numbers for a planet with a core of
constant mass, $M_c = 5M_\oplus$, as a function of the core shear
modulus $\mu$, for three different core radii. For the range of core
sizes considered (up to $R_c/R = 0.15$), changing the shear modulus by
$4$ orders of magnitude apparently hardly changes the surface tidal
$k_2$ (bottom panel). Core radius plays a slightly more important
role, with smaller cores yielding smaller $k_2$, as expected (note
this is opposite to Fig. 1 because here we are keeping core mass
rather than core density constant). Perhaps surprisingly, smaller
cores are deformed more than larger cores (top panel). This can be
understood by noting that the amount of deformation depends on
$\bar{\mu}$, the ratio of the core shear modulus to the the
gravitational rigidity $\rho_c g_c R_c$ of the core, with larger
ratios yielding smaller deformations (cf. Eq. \ref{h0k0}). For
constant core mass, this ratio scales as $\bar{\mu} \propto R_c^4$,
and therefore the core deforms less at higher radii.

\begin{figure}
\scalebox{0.58}{\includegraphics{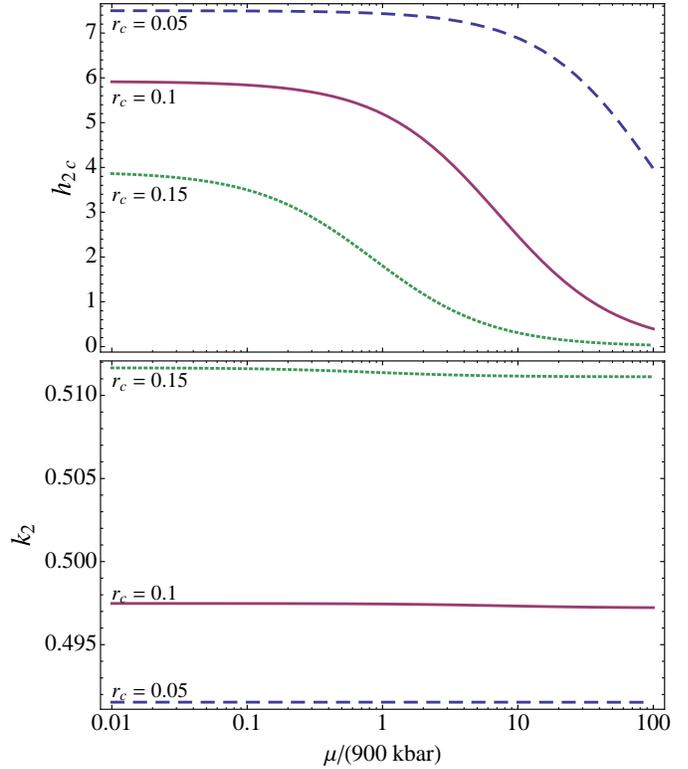}}
\caption{Sample tidal Love numbers $h_{2c}$ (top) and $k_2$ (bottom)  for a gas giant with mass $M_J$, radius $R_J$, and core mass $M_c=5 M_\oplus$, as a function of shear modulus $\mu$ for several values of fractional core radius $r_c \equiv R_c/R$.}
\end{figure}

\subsection{Dissipation of viscoelastic core}

We now consider tidal dissipation of a viscoelastic core, modeled by a complex shear modulus. We take
\be
\mu \rightarrow \tilde{\mu} \equiv \tilde{\mu}_1 + i \tilde{\mu}_2,
\ee
and assume the simplest viscoelastic model - the Maxwell model, such that (Henning et al. 2009)
\ba
\tilde{\mu}_1 &=& \frac{\mu \left(\omega/\omega_M\right)^2}{1+\left(\omega/\omega_M\right)^2}, \\
\tilde{\mu}_2 &=& -\frac{\mu \left(\omega/\omega_M\right)}{1+\left(\omega/\omega_M\right)^2}, 
\ea
where $\omega$ is the forcing frequency in the reference frame of the
planet, and $\omega_M$ is the Maxwell frequency given by
$\omega_M\equiv\mu/\eta$, where $\mu$ is the normal shear modulus of
the core and $\eta$ the viscosity of the core. Under the Maxwell
model, the solid core responds viscously for $\omega \lo \omega_M$ and
elastically for $\omega \go \omega_M$.

Figure 3 shows tidal dissipation rates, characterized by 
${\rm Im}[k_2]$, as a function of the forcing frequency, for different
values of $\mu$ and $\eta$. Since $\eta$ only enters into the
expression for $\mu$ through the ratio $\omega/\omega_M$, it is not
surprising that changing $\eta$ simply shifts the tidal dissipation
curve horizontally without changing the strength (Fig. 3, bottom). The
effect of $\mu$ is more complicated (Fig. 3, top panel) and can shift
the curve up/down as well. Since $\mu$ does not directly affect the
viscous properties of the core, it makes sense that a change in $\mu$
shifts the curve such that the tidal response on the viscous side
($\omega \lo \omega_M$) remains unchanged.

\begin{figure}
\scalebox{0.58}{\includegraphics{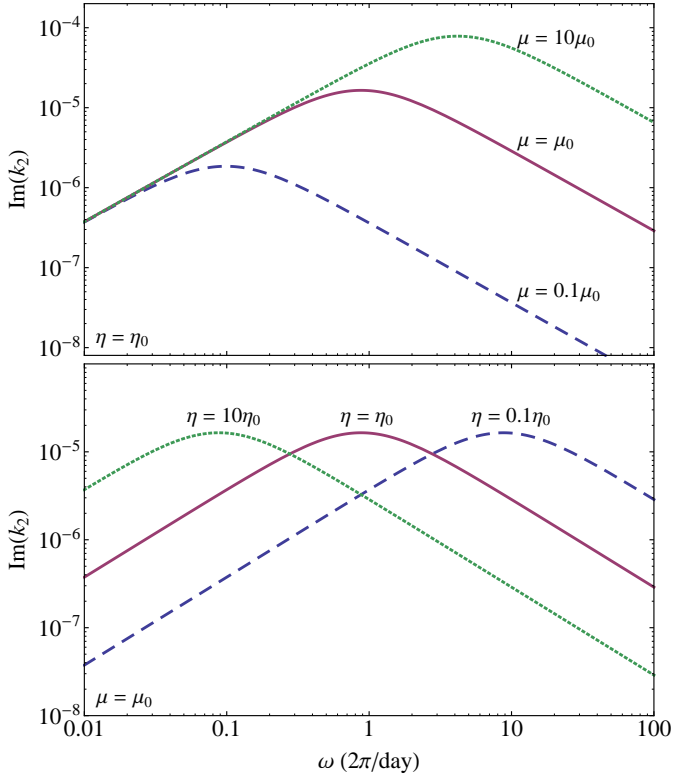}}
\caption{Sample tidal dissipation rates for a gas giant with mass $M_J$, radius $R_J$ with an $n=1$ envelope, characterized by ${\rm Im}[k_2]$, as a function of the forcing frequency $\omega$. Here $R_c = 0.1 R$, and $M_c = 5M_\oplus$.Top panel: for a fixed viscosity $\eta=\eta_0\approx12.3\,\rm{Gbar \cdot s}$ and three values of $\mu$, were $\mu_0 = 900$\,kbar. Bottom panel: for a fixed $\mu=\mu_0$ and three values of $\eta$.}
\end{figure}

\subsection{Comparison with planet models with uniform-density envelope}

An analytical formula for the tidal number $k_2$ was previously
derived for giant planet models with uniform envelope density (Dermott
1979).  Recent works have used viscoelastic dissipation in the solid
cores of such models to explain the amount of tidal dissipation
inferred from the evolution of Jupiter's and Saturn's satellites
(Remus et al. 2012, Remus et al. 2015) and from constraints on
high-eccentricity migration of hot Jupiters (Storch \& Lai 2014). In
Figure 4 we compare the dissipation levels in planets with $n=1$ vs
uniform-density envelopes. 
While we do not attempt to explore the full parameter space here,
Figure 4 suggests that the
difference between the two depends most strongly on the density of the
core, with the $n=1$ dissipation being stronger in more compact cores
by as much as a factor of few.

\begin{figure}
\scalebox{0.58}{\includegraphics{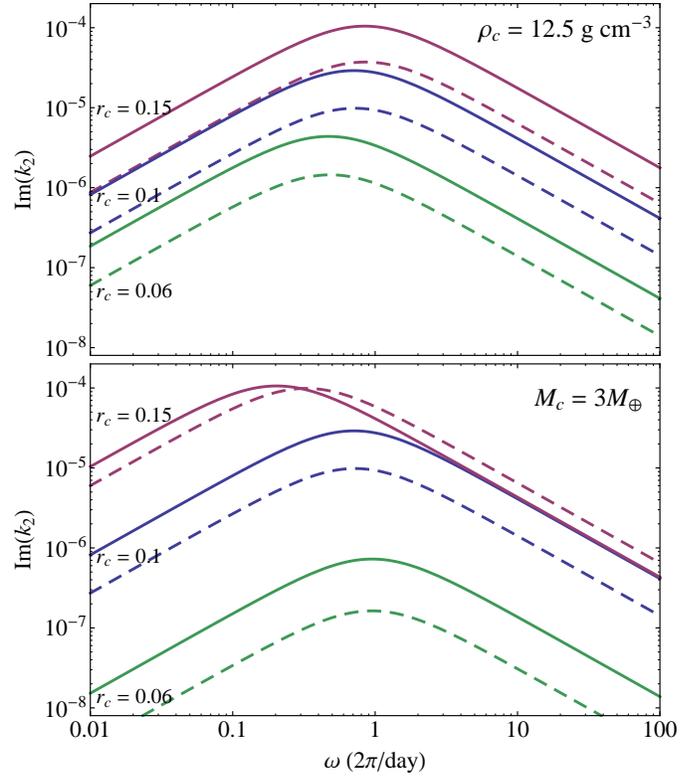}}
\caption{Comparison of tidal response curves for a gas giant with mass $M_J$, radius $R_J$ with an $n=1$ envelope (solid lines) vs $n=0$ envelope (dashed lines). Top panel: for a solid core of fixed density $\rho_c = 12.5{\rm g~cm^{-3}}$ but varying radii $r_c\equiv R_c/R$. Bottom panel: for a solid core of fixed mass $M_c = 3M_\oplus$ but varying radii. The core mass and density in each panel have been selected such that the $r=0.1$ curves are identical in the top and bottom panels. We assume $\mu=\mu_0=900$\,kbar and $\eta=\eta_0=12\,{\rm Gbar\cdot s}$.}
\end{figure}

\subsection{Application to Super-Earths}

While in this work we focus mainly on gas giants, the formulae
presented in section 3 do not assume the core radius to be
small. Thus, in principle, they may be applied to super-Earths - with
the caveat that super-Earths are not likely to be well-described by
$n=1$ envelopes. In Figure 5 we present the Love numbers for a
super-Earth analogue similar to Kepler-11d (Lissauer et al. 2013),
consisting of a solid core and a $n=1$ gas envelope that is $\sim15\%$
by mass, but $\sim50\%$ by radius. In this case the effect of the
``core'' shear modulus on the surface $k_2$ is significant when $\mu$
is varied by a few orders of magnitude.

\begin{figure}
\scalebox{0.58}{\includegraphics{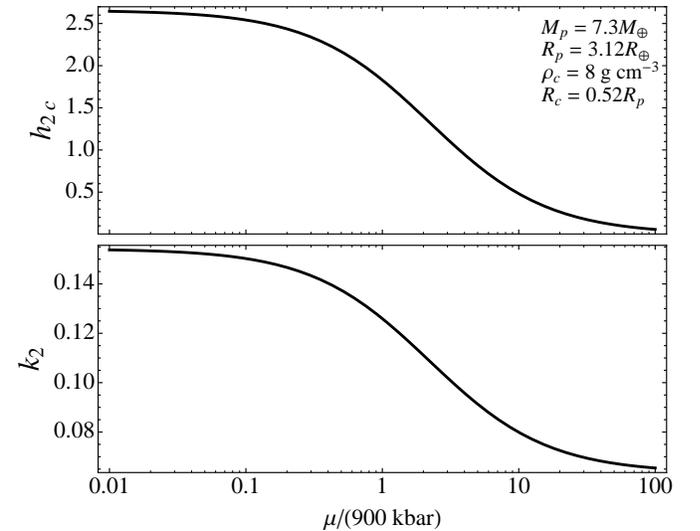}}
\caption{Tidal response for a super-Earth analogue similar to Kepler-11d, assuming an $n=1$ envelope.}
\end{figure}

%%%%%%%%%%%%%%%%%%%%%%%%%%%%%%%%%%%%
\section{Conclusion}

We have presented a general method of computing the tidal Love numbers
of giant planets consisting of a uniform elastic solid core and a
non-uniform fluid envelope. We show that if the envelope obeys the
$n=1$ polytropic equation of state ($P\propto \rho^2$), 
simple analytical expressions for the tidal Love numbers can be
obtained [see Eqs.~(\ref{h2c})-(\ref{lambda})].
These expressions are valid for any core size, density, and shear
modulus.  They allow us to compute the tidal dissipation rate in the
viscoelastic core of a planet by using a complex shear modulus that
characterizes the rheology of the solid core.  Our results improve
upon previous works that are based on planetary models with a solid core
and a uniform ($n=0$) envelope. 
In general, we we find that while for diffuse (low density, larger
size) cores, the dissipation rates of the $n=0$ envelope models can be
higher than the $n=1$ models, for more compact cores the $n=1$
envelope models have higher dissipation rates by as much as a factor
of few.

While we have focused on analytical expressions for the Love numbers
in this paper, our method and equations can be adapted for numerical
computation of the real (non-dissipative) Love numbers
for any envelope equation of state. In particular, they can be used
to study tidal distortion of super-earths, which typically contain
a H-He envelope (a few percent by mass) surrounding a rocky core (e.g.,
Lissauer, Dawson \& Tremaine 2014), or tidal distortion in gas giants with more
realistic equations of state.
It is possible that for some exoplanetary systems, particularly those containing hot Jupiters,
the tidal Love number $k_2$ can be constrained or measured
using secular planet-planet interactions (e.g. Mardling 2010, Batygin \& Becker 2013).
This would provide a useful probe of the interior structure of the planet.

In the presence of viscosity in the solid core, additional
work is still needed to obtain the tidal viscoelastic dissipation
rate even when numerical results for the real Love numbers are
available. Thus, our analytical expressions will be useful, as they
allow for simple computation of the tidal dissipation rate via the
correspondence principle, and can serve as a calibration of
numerical results.

\section*{Acknowledgments}
This work has been supported in part by NSF grant
AST-1211061, and NASA grants NNX14AG94G and NNX14AP31G.

%%%%%%%%%%%%%%%%%%%%%%%%%%%%%%%%%%%%%%%%%%%%%%%%%

\end{document}